\documentclass[10pt, conference, letterpaper]{IEEEtran}
\usepackage{graphicx}
\usepackage{xcolor}
\usepackage{amsmath}
\usepackage{tikz}
\usepackage{pgf-umlsd}
\usetikzlibrary{shapes,arrows,shadows} 
\usepackage{amsthm}
\usepackage{bm}
\usepackage{tikz}
\usepackage{subcaption}
\usepackage{amssymb}
\usepackage[font=small,labelfont=bf]{caption}
\usepackage{algorithm}
\usepackage{algpseudocode}
\usepackage{pgf}
\usetikzlibrary{arrows,automata,shadows}
\usepackage[latin1]{inputenc}
\usetikzlibrary{decorations.pathreplacing}

\usepackage[left=0.65in,right=0.65in,top=0.75in,bottom=1in]{geometry}
\DeclareMathOperator*{\expectation}{E}

\graphicspath{{./images/}}

\usepackage[absolute,showboxes]{textpos}

\TPMargin{5pt}

 \theoremstyle{definition}

\definecolor{TUMBlau}{RGB}{0,101,189} 
\definecolor{TUMBlauDunkel}{RGB}{0,82,147} 
\definecolor{TUMBlauHell}{RGB}{152,198,234} 
\definecolor{TUMBlauMittel}{RGB}{100,160,200} 

\definecolor{TUMElfenbein}{RGB}{218,215,203} 
\definecolor{TUMGruen}{RGB}{162,173,0} 
\definecolor{TUMOrange}{RGB}{227,114,34} 
\definecolor{TUMGrau}{gray}{0.6} 

\newcommand{\nactiveuserrv}{N_{\text{a}}} 
\newcommand{\nsucuserresource}{N_{\text{s}}} 
\newcommand{\nactiveuser}{n_{\text{a}}} 
\newcommand{\totaluser}{n_\text{tot}} 
\newcommand{\device}{user} 
\newcommand{\parresource}{g} 
\newcommand{\latency}{\mathrm{L}} 
\newcommand{\numberresolved}{N_\text{r}} 
\newcommand{\mpr}{K} 
\newcommand{\simprob}{\pi} 

\newcommand{\reliability}{\mathrm{R} (\latency)} 
 
\newcommand{\gstar}{{\parresource}^*} 

\definecolor{gl}{rgb}{0.0,0.5,0.8}
\definecolor{fc}{rgb}{0.8,0.5,0}
\definecolor{al}{rgb}{1,0.3,0.3}

\begin{document}

%

\title{On Throughput Maximization of Grant-Free Access with Reliability-Latency Constraints}

\author{\IEEEauthorblockN{H. Murat G\"ursu\IEEEauthorrefmark{1}, Wolfgang Kellerer\IEEEauthorrefmark{1}, \v Cedomir Stefanovi\' c\IEEEauthorrefmark{2}}
\IEEEauthorblockA{\IEEEauthorrefmark{1}Chair of Communication Networks, Technical University of Munich, Munich, Germany\\
\IEEEauthorrefmark{2}Department of Electronic Systems, Aalborg University, Aalborg, Denmark \\
E-mail:\{murat.guersu,wolfgang.kellerer\}@tum.de, cs@es.aau.dk}}

\maketitle

%

\begin{abstract}
Enabling autonomous driving and industrial automation with wireless networks poses many challenges, which are typically abstracted through reliability and latency requirements. One of the main contributors to latency in cellular networks is the reservation-based access, which involves lengthy and resource-inefficient signaling exchanges.
An alternative is to use grant-free access, in which there is no resource reservation.
A handful of recent works investigated how to fulfill reliability and latency requirements with different flavors of grant-free solutions.
However, the resource efficiency, i.e., the throughput, has been only the secondary focus. 
In this work, we formulate the throughput of grant-free access under reliability-latency constraints, when the actual number of arrived users or only the arrival distribution are known.
We investigate how these different levels of knowledge about the arrival process influence throughput performance of framed slotted ALOHA with $K$-multipacket reception, for the Poisson and Beta arrivals.
We show that the throughput under reliability-latency requirements can be significantly improved for the higher expected load of the access network, if the actual number of arrived users is known.
This insight motivates the use of techniques for the estimation of the number of arrived users, as this knowledge is not readily available in grant-free access.
We also asses the impact of estimation error, showing that for high reliability-latency requirements the gains in throughput are still considerable.  

\end{abstract}
\vspace{-0.1cm}
\vspace{-0.1cm}

\section{Introduction}

One of the key novelties of 5G research, development and standardization is that it explicitly addresses reliability and latency requirements.
This is best exemplified in the introduction of a novel service category -- Utra Reliable and Low Latency Communications (URLLC), for which the generic requirement is the reliability of $1 - 10^{-5}$ (i.e., 0.99999) with the user-plane radio-latency of 1~ms for a single transmission of a 32-byte long packet.
Nevertheless, use cases from the other foreseen 5G service categories, which are enhanced Mobile BroadBand (eMBB) and massive Internet-of-Things (mIoT), also involve reliability and latency requirements~\cite{D2.1}.
For instance, smart-cities use-case, which belongs to mIoT, involves reliability of 0.95 with user-plane radio-latency of 0.5~ms~\cite{D2.1}.

On the other hand, cellular networks are characterized by reservation-based access, which involves a random-access based, signaling-intensive connection-establishment procedure~\cite{tyagi2017connection}.
This approach is highly inefficient when short packets are sporadically exchanged, which is characteristic for IoT use-cases~\cite{laya2014random,MNKPSP2016}.
Moreover, each stage of the connection-establishment has the potential to compromise reliability and increase latency~\cite{ericson2018initial,PSNCATB2018}.
Thus, 3GPP has decided to standardize a grant-free access method, alongside the existing resource-reservation, in which the users will contend with their data packets in random-access fashion~\cite{R1-1808304}.

Besides fulfilling reliability and latency requirements, a grant-free access scheme should also maximize the efficiency of the use of the time-frequency resources dedicated to it, i.e., maximize the \emph{throughput}.
However, the aspect of throughput maximization has been less investigated in the existing works on grant-free access.
Nevertheless, efficient use of resources plays an important role in the overall framework of radio-resource management in cellular access networking.


In this paper, we investigate the throughput maximization of the grant-free access from the perspective of medium-access control.
Specifically, we analyze how the knowledge of number of arrived (i.e., contending) users can be used to boost the throughput, providing the following contributions:
\begin{itemize}
\item  We give a formal definition of reliability under predefined latency constraint for the batch arrival, developing it for the cases when the number of arrived users is exactly known, or given by a certain arrival distribution. We also formally define throughput for both cases, providing insight on the role of the knowledge of the number of arrived users.
\item We derive throughput under the reliability-latency requirements for framed slotted ALOHA (FSA) with $K$-multipacket reception (MPR), i.e., we assume that the operation of physical layer can be represented with successful reception up to and including $K$ packets that occur simultaneously in a slot.
\item We instantiate the analysis for the cases of Poisson and Beta arrivals, which are standard models of IoT traffic, and evaluate the impact of $K$-MPR and the knowledge of the number of arrived users. We show that increasing $\mpr$ as well as the knowledge of the number of arrived users pay off in throughput.
\item The last insights suggests that estimation of the number active users in grant-free access can be beneficial, as this information is typically not readily available. In this respect, we investigate the impact of the potential estimation errors on the throughput, showing that for high reliability-latency requirements the gains are still considerable, even with high error levels.
\end{itemize}


The rest of the text is organized as follows.
A brief overview of the related work is made in Section~\ref{sec:sota}.
Section~\ref{sec:sys_mod} introduces the system model. 
Section~\ref{sec:params} defines the analytical framework composed of reliability-latency requirement and throughput maximization, which is then applied to FSA with $\mpr$-MPR in Section~\ref{sec:rrm}.
Section~\ref{sec:eval} evaluates the performance of FSA with $\mpr$-MPR under Poisson and Beta arrivals and the effect of user activity estimation. 
Section~\ref{sec:conc} concludes the paper.



\section{Related Work}
\label{sec:sota}


Grant-free access from the system level perspective was investigated in \cite{jacobsen2017system}, for an outdoor 3GPP urban scenario with 21 cells.
The main contribution of the paper are simulation-based results that outline the setups in which the grant-free approaches outperform the grant-based ones, for the case of Poisson arrivals.
Another work evaluates user activity with Bernoulli arrivals with different activation probabilities for grant-free scenarios in \cite{kotaba2018uplink}, proposing a new hybrid scheme that benefits from the advantages of both grant-based and grant-free schemes and focusing on the achievable data rates.
A system level integration of grant free $\mpr$-MPR in the 5G setup is evaluated via simulations in \cite{saur2017radio}, comparing the effects of the channel estimation failure and contention failure on the access protocol design.

Another line of works considers advanced random access algorithms from reliability and/or latency perspective.
Work~\cite{jian2017random} derives delay distribution for the multichannel slotted ALOHA.
In \cite{singh2018contention}, $\mpr$-MPR FSA is evaluated and the authors provide analytical expressions for collision probability under Poisson arrivals; we note that the extension for other arrival types is not trivial.
Moreover, the resource-efficiency perspective is neglected, as the throughput is not evaluated, and there is no discussion on how increasing $\mpr$ affects the throughput.
An extension of Irregular Repetition Slotted ALOHA (a slotted ALOHA-based scheme with successive interference cancellation) for the scenarios with multiple classes of Beta arrivals with different reliability-latency constraints is investigated in~\cite{polling_abbas}. 
Frameless ALOHA (another scheme that also exploits successive interference cancellation) with reliability-latency guarantees is considered in \cite{stefanovic2017frameless}.
Finally, tree algorithms with multiple channels for reliability-latency constraints are analyzed in~\cite{gursu2017delay}.


\vspace{-0.1cm}
\section{System Model}
\label{sec:sys_mod}

\begin{figure}[t!]
	\centering
	\includegraphics[width=0.33\textwidth]{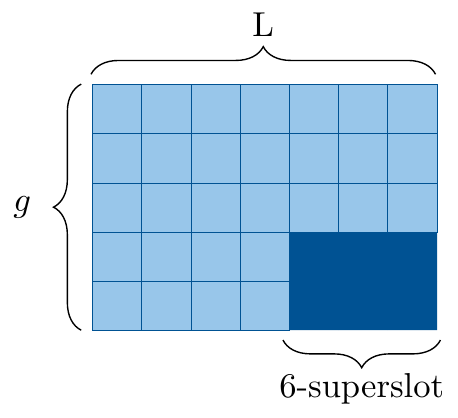}
	\caption{The resource grid comprising $\latency$ time slots and $\parresource$ channels; $\latency$ is given by the latency budget and $\parresource$ is optimized such that target performance is achieved. The figure also shows a 6-superslot defined over 2 channels and 3 time slots.}	\label{fig:grantfree_rg}
	\vspace{-0.1cm}
	\vspace{-0.1cm}
	\vspace{-0.1cm}
\end{figure}

We focus on a single cell with a homogenous population of $\totaluser$ users that access the common access point (AP). 
The users are randomly and sporadically activated, and their activity is modeled via a batch arrival of $\nactiveuserrv$ users. 
In general, $\totaluser$ (or some upper bound on it) is assumed to be known, while $\nactiveuserrv$ is a random variable.
The time-frequency resources in the uplink are divided in a grid consisting of time-frequency slots (denoted simply as slots in further text), and without loss of generality, we assume that the slot-bandwidth and slot-duration are of a unit size.
Fig.~\ref{fig:grantfree_rg} shows the model of the resource grid. 

Further, we assume that the available slots are grouped in $K$-\emph{superslots}: a $K$-superslot is dimensioned such that if there are up to and including $K$ simultaneous transmissions occurring in it, all of them are successfully received (and the corresponding users become resolved).
Otherwise, if there are more than $K$ transmissions occurring in a $K$-superslot, none of them can be successfully received.
In other words, we assume that the physical layer operation can be represented by $K$-MPR. The signaling overhead required to obtain the channel estimation needed to enable K-MPR is evaluated in \cite{gursu2018multiplicity} and in this work we assume the channel state information is available. We also assume that a $K$-superslot contains $K$ slots in order to achieve the $K$-MPR capability and note that the linear increase in the superslot size with $K$ is a reasonable assumption, cf.~\cite{GSP2018,MGA2017}. Finally, we assume that the users are aware of the superslot boundaries.
This type of synchronization could be achieved via means of a downlink control channel, which is the typical scenario in cellular systems.
Fig.~\ref{fig:grantfree_rg} shows an example of 6-superslot.

For $K=1$, the above model reduces to the standard collision channel model.
Moreover, although simplistic, this model of $K$-MPR can be used as an approximation for systems in which other sources of diversity are employed to achieve multipacket reception, like the use of spreading codes, or multiple antennas.

The access decision of users is regulated via a grant-free access algorithm, whose goal is to ensure a predefined level of reliability $\reliability$ of user resolution under a predefined latency constraint $\latency$ in time units, see Fig.~\ref{fig:grantfree_rg}.
We denote this requirement as the \emph{reliability-latency} requirement in further text. 
Note that in the proposed setup, the number of frequency channels $\parresource$ assigned to the access procedure is the degree of freedom that can be optimized such that the target reliability $\reliability$ is achieved. This reflects a typical radio resource management problem.

\vspace{-0.1cm}
\vspace{-0.1cm}
\vspace{-0.1cm}
\section{Performance Parameters}
\label{sec:params}
\vspace{-0.1cm}
\subsection{Reliability-latency}
\vspace{-0.1cm}
Formally, denote by $\xi_u$ the event that an active user $u$ becomes resolved and by $\latency $ the maximum allowed latency of the resolution in time-units.
In case of batch arrivals, the access algorithm should satisfy the following reliability-latency definition
\vspace{-0.1cm}
\begin{align}
\label{eq:rl_basic}
r ( \parresource, \latency, \nactiveuser ) = \Pr [ \xi_u, L \leq \latency | \parresource, \nactiveuser, \totaluser ] \geq \reliability
\end{align}
for all active users $u$ in the batch, where it is assumed that the realization $\nactiveuser$ of $\nactiveuserrv$ is known.
Assuming that the access algorithm does not output false positives in  (i.e., an inactive user can not be falsely resolved as active), the above condition can be expressed as
\begin{align}
r ( \parresource, \latency, \nactiveuser ) &  =  \sum_{k =1}^{\nactiveuser} \frac{k}{ \nactiveuser}  \Pr [ \numberresolved = k, L \leq \latency | \parresource,\nactiveuser, \totaluser ]  \nonumber \\ 
\label{eq:rl} 
 & =   \frac { \mathrm{E} [ \numberresolved ] }{\nactiveuser} \geq \reliability
\end{align}
where $\numberresolved$ is the number of resolved users, and $\frac{k}{ \nactiveuser}$ is the probability that active user $u$ is among the $k$ resolved ones.
In the assumed system model, $ g = g ( \nactiveuser)$ should be chosen such that the condition \eqref{eq:rl} becomes satisfied.

If the realization $\nactiveuser$ is not known, the use of condition \eqref{eq:rl} is not possible.
However, if the probability mass function (pmf) of $\nactiveuserrv$ is known, the reliability-latency condition could be defined as:
\vspace{-0.1cm}
\vspace{-0.1cm}
\begin{align}
\label{eq:rl_total}
r^*( \parresource^*, \latency ) = \sum_{\nactiveuser} r ( \parresource^*, \latency, \nactiveuser ) \Pr [ \nactiveuserrv = \nactiveuser ] \geq \reliability
\end{align}
where $\parresource^*$ should be chosen such that condition \eqref{eq:rl_total} is satisfied.


It is natural to assume that for any reasonable access algorithm, the following holds:
\begin{align}
r ( \parresource, \latency, n ) \leq r ( \parresource + 1, \latency, n ), \, \forall n
\end{align}
i.e., increasing the number of frequencies (which increases the total number of resources) will not lower chances to fulfill the reliability-latency condition.
Under this assumption, it could be shown that there are minimal values for $\parresource (\nactiveuser)$ and $\parresource^*$, for which \eqref{eq:rl} and \eqref{eq:rl_total}, respectively, hold. 
Along the same lines, one can formulate optimization problems according to which these minimal values can be found, respectively
\begin{align}
\parresource_\text{min} (\nactiveuser) & = \arg\min_\parresource (r( \parresource, \latency, {\nactiveuser} ) : r( \parresource, \latency, {\nactiveuser} ) = \reliability ) \label{eq:gopt}\\
\parresource_\text{min}^* & = \arg\min_\parresource \left( r^*( \parresource, \latency) : r^* ( \parresource, \latency ) = \reliability \right ) \label{eq:gstar}.
\end{align}
For the sake of brevity and with a slight abuse of notation, in the rest of the text we will assume that $\parresource (\nactiveuser) = \parresource_\text{min} (\nactiveuser) $ and $\parresource^* = \parresource_\text{min}^*$, respectively.
\vspace{-0.1cm}
\subsection{Throughput}

The number of resources dedicated to the resolution is $\parresource \, \latency$, see Fig.~\ref{fig:grantfree_rg}. 
We define the throughput as the expected number of resolved users vs. the number of resources
\begin{align}
\mathrm{T} = \mathrm{E} \left[ \frac{\numberresolved}{ \parresource \latency}   \right ].
\end{align}

In the proposed model, $\numberresolved$ is determined by the employed reliability-latency condition and the throughput is maximized by minimizing $\parresource$.
Specifically, when \eqref{eq:rl} is used, the throughput becomes
\vspace{-0.1cm}
\begin{align}
\mathrm{T} = \frac{\reliability}{\latency} \sum_{\nactiveuser} \frac{ \nactiveuser } { \parresource (\nactiveuser)  } \Pr [ \nactiveuserrv = \nactiveuser ] ,
\label{eq:t_g}
\end{align}
where we recall that $  \parresource (\nactiveuser)  $ is chosen such that $r ( \parresource, \latency, \nactiveuser  ) = \reliability$, $\forall\,\, \nactiveuser $.
In case when \eqref{eq:rl_total} is used, the throughput is
\begin{align}
\mathrm{T}^* = \frac{\reliability}{\parresource^* \latency} \sum_{\nactiveuser} \nactiveuser \Pr [ \nactiveuserrv = \nactiveuser ] = \frac{\reliability \, \mathrm{E} [ \nactiveuserrv ] }{\parresource^* \latency} .
\label{eq:t_g*}
\end{align}
where we also recall that $\parresource^*$ is chosen such that $r^* ( \parresource^*, \latency ) = \reliability$.
Obviously, there is a difference between $\mathrm{T}$ and $\mathrm{T}^*$, which will be further investigated in Section~\ref{sec:eval}. 


\section{Grant-Free Access with FSA}
\label{sec:rrm}

\begin{figure}
		\centering
	\includegraphics[width = 0.5\textwidth]{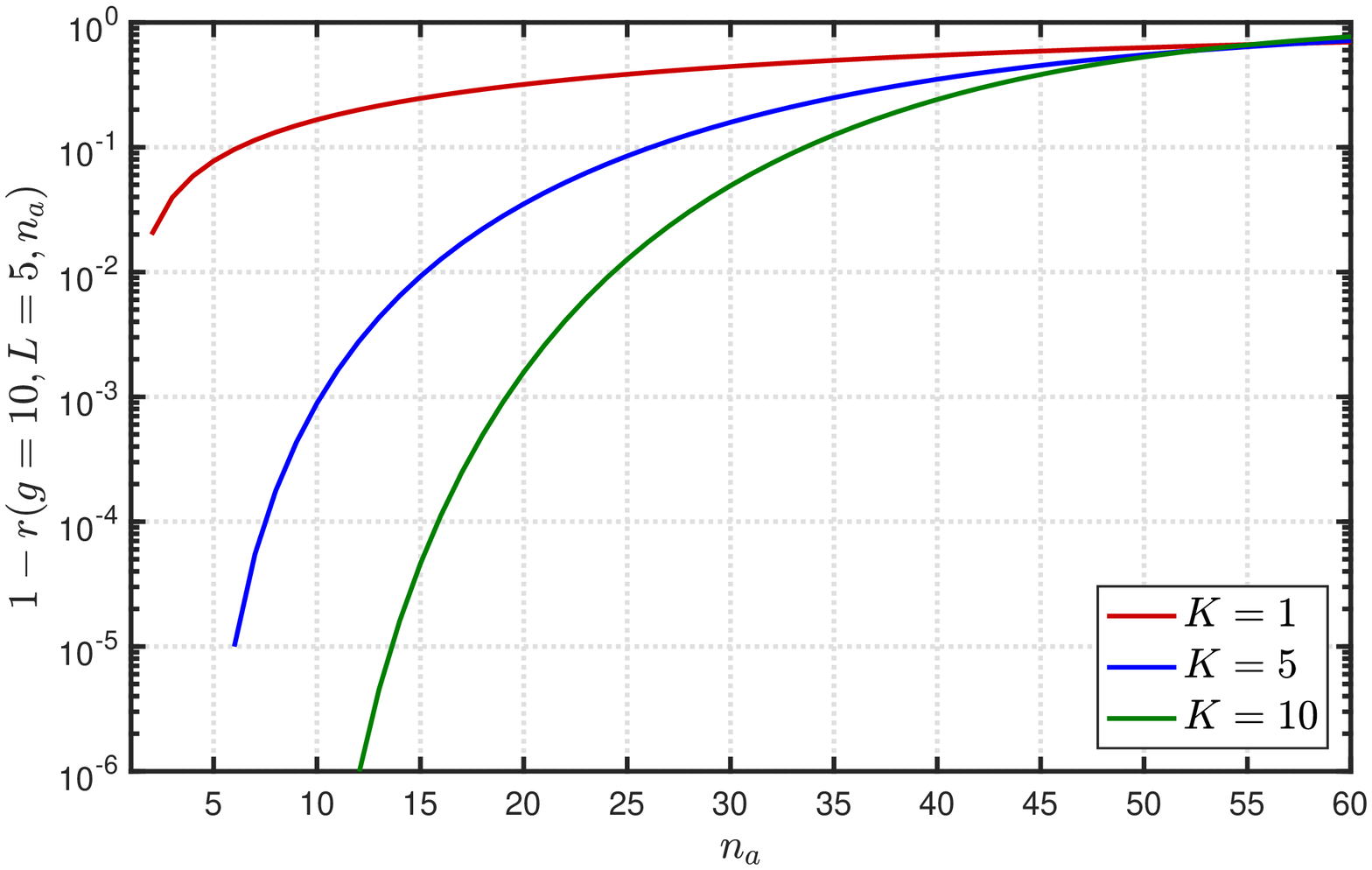}
	\caption{Reliability-latency performance of framed slotted ALOHA with $K$-MPR for varying number of \device s $\nactiveuser$~and $\mpr$, when $\parresource= 10$ and $\latency = 5$.}
	\label{fig:rel_kmpr}
	\vspace{-0.1cm}
	\vspace{-0.1cm}
\end{figure}


In the considered scenario, the frame consists of $ \parresource \, \latency$ slots grouped in $K$-superslots. Thus, there are  $\lfloor\frac{\parresource \, \latency}{K}\rfloor$ $K$-superslots.
In FSA, each of the active \device s transmits its packet in a uniformly randomly chosen $K$-superslot of the frame.\footnote{A related analysis to the one considered here is made in \cite{singh2018contention}, where the authors derived only $\mathrm{T}^*$ for the case when $K$ is fixed to 8, but there is neither investigation of the behavior of $\mathrm{T}$, nor the impact of varying $K$.}  

For the given $\nactiveuser$, $\latency$, $\parresource$ and $K$, the reliability of FSA can be calculated as
\begin{equation}
r  ( \parresource, \latency, \nactiveuser ) = \left({1-\simprob}\right)^{\nactiveuser-1} \sum_{i=0}^{\mpr-1} {{\nactiveuser-1}\choose{i}} \left(\frac{\simprob}{1-\simprob}\right)^{i}
\label{eq:def_rel_mcsa_mpr2}
\end{equation}
where $\simprob = \frac{1}{\lfloor\frac{\parresource \, \latency}{K}\rfloor}$ is the probability of choosing a certain $K$-superslot.
The proof of \eqref{eq:def_rel_mcsa_mpr2} is given in Appendix~\ref{app:2}.
Using \eqref{eq:def_rel_mcsa_mpr2}, as well as substituting it into \eqref{eq:rl_total}, one can find the values of $\parresource (\nactiveuser)$ and $\parresource^*$ through \eqref{eq:gopt} and \eqref{eq:gstar}, as well as of $\mathrm{T}$ and $\mathrm{T^*}$ through \eqref{eq:t_g} and \eqref{eq:t_g*}, respectively.


In order to illustrate the effect of increasing $K$ on reliability-latency performance, we consider an example where $\parresource=10$ and $\latency=5$, i.e., there are $\parresource \, \latency = 50$ slots available.
Fig.~\ref{fig:rel_kmpr} shows how $ r ( \parresource, \latency, \nactiveuser)$ behaves when $\nactiveuser$ is varied in such setup.
Obviously, for $\nactiveuser \leq 50$, increasing $K$ has a beneficial effect on $ r ( \parresource, \latency, \nactiveuser)$; such trend would continue until $K$ reaches 50, when $ r ( \parresource, \latency, \nactiveuser)$ would become 1. 
On the other hand, the optimal $K$ for $\nactiveuser > 50$ depends on $\nactiveuser$.
Nevertheless, note that for $\nactiveuser > \parresource \, \latency$, the achievable levels of  $ r ( \parresource, \latency, \nactiveuser)$ are small, as there are more arrived users than the number of slots.

In the next section, we turn to the throughput maximization for FSA with $\mpr$-MPR when the reliability-latency requirement is fixed to $\reliability$.

\begin{figure*}[t]
\centering
\begin{subfigure}{\columnwidth}
		\centering
	\includegraphics[width = 1\textwidth]{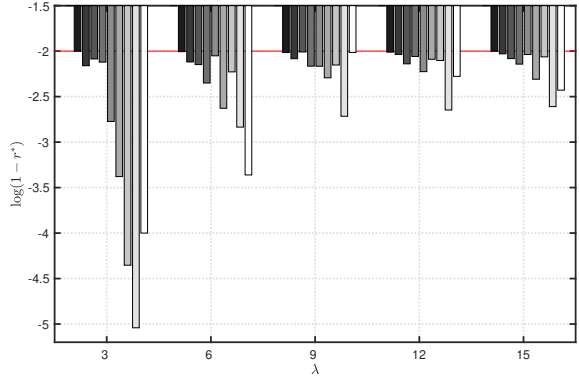}
	\caption{Poisson arrivals }
	\label{fig:gstar_09}
\end{subfigure}
\begin{subfigure}{\columnwidth}
	\includegraphics[width = 1\textwidth]{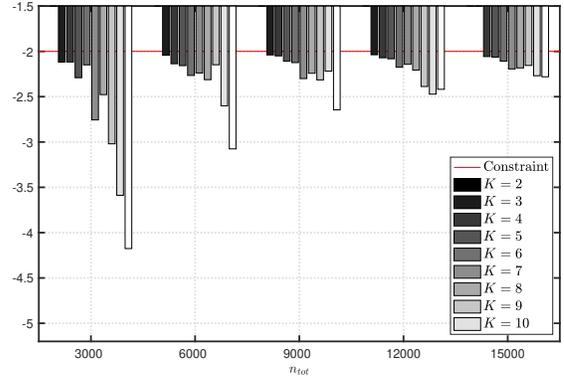}
	\caption{Beta arrivals}
	\label{fig:gstar_099}
\end{subfigure}
\caption{{ $1- {r^* ( \gstar, \latency} )$ for different values of $\mpr$, and (a) Poisson and (b) Beta arrivals, when $\latency=5$ and $ \reliability = 0.99$.}}
\label{fig:gstar_rel_sims}
\vspace{-0.1cm}
\vspace{-0.1cm}
\end{figure*}

\begin{figure*}[t]
	\centering
	\begin{subfigure}{0.66\columnwidth}
		\includegraphics[width = 1\textwidth]{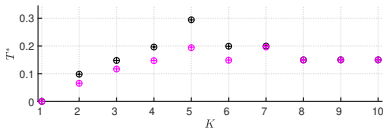}
		\caption{$\lambda=3$ and $\totaluser=3000$}
		\label{fig:gstar_091}
	\end{subfigure}
	\begin{subfigure}{0.66\columnwidth}
		\includegraphics[width = 1\textwidth]{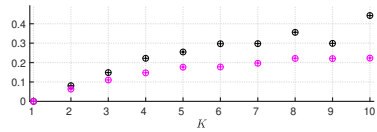}
		\caption{ $\lambda=9$ and $\totaluser = 9000$}
		\label{fig:gstar_0992}
	\end{subfigure}
	\begin{subfigure}{0.66\columnwidth}
		\includegraphics[width = 1\textwidth]{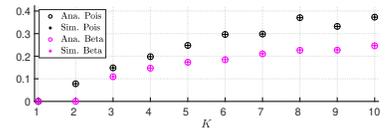}
		\caption{$\lambda=15$ and $\totaluser=15000$}
		\label{fig:gstar_0999}
	\end{subfigure}
	\caption{$\mathrm{T}^*$ for different values of $\mpr$ and Poisson and Beta arrivals, when $\latency=5$ and $ \reliability = 0.99$; the subplots depict results obtained for the same expected number of arrived users per frame.}
	\label{fig:gstar_tpt_sims}
	\vspace{-0.1cm}
	\vspace{-0.1cm}
\end{figure*}

\vspace{-0.1cm}
\vspace{-0.1cm}
\vspace{-0.1cm}
\section{Evaluation}
\label{sec:eval}

In this section, we compare the throughputs $\mathrm{T}$ and $\mathrm{T}^*$, given by {\eqref{eq:t_g}} and {\eqref{eq:t_g*}}, respectively, for varying $\mpr$ and fixed $\latency$.
The arrival distribution of $\nactiveuserrv$ is modeled by (i) a Poisson and (ii) a {Beta} distribution.
For the Poisson distribution, we have
\begin{equation}
	\Pr [ \nactiveuserrv = \nactiveuser |  \lambda ]  = \frac{\lambda^{\nactiveuser} }{\nactiveuser !} e^{-\lambda}
\end{equation}
where the mean number of arrived users is $\mathrm{E} [ \nactiveuserrv]  = \lambda$, assumed to be known.

According to the beta distribution, the probability that a device is activated in time instant $t \in [ 0, T_A]$ is given by  
\begin{align}
p (t) = \frac{t^{\alpha-1} (T_A - t)^{\beta -1 }}{T_A^{\alpha+\beta -1} \textrm{Beta}(\alpha,\beta)}
\end{align}
where $\textrm{Beta}(\alpha,\beta)= \int^1_0 t^{\alpha-1} (1-t)^{\beta-1}dt$, $\alpha$ and $\beta$ are shape parameters, and $T_A$ is the activation time.
We assume the 3GPP model~\cite{3rdGenerationPartnershipProject3GPP}, where $\alpha=3$, $\beta=4$ and $T_A = 10$~s.
Assuming that the activation time is discretized into intervals of length $\latency$ time units\footnote{I.e., we assume that the arrivals are gated in batches of $\latency$ time units.}, the probability of $\nactiveuser$ arrivals in interval $t_s$, ${t_s \in [0,\cdots, \frac{T_A}{\latency} -1]}$, is
 \begin{equation}
 \label{eq:b}
 \Pr [ \nactiveuserrv = \nactiveuser | t_s]  = {{\totaluser}\choose{\nactiveuser}} P [t_s]^{\nactiveuser}(1 - P [t_s])^{\totaluser - \nactiveuser}
 \end{equation}
where the total number of $\totaluser$ users is assumed known and  $P [t_s]$ is given by 
\vspace{-0.1cm}
 \begin{equation}
 P [t_s] = \int\limits_{t_s \latency}^{(t_s + 1) \latency} p (t) \, dt.
\end{equation}
This approach is reminiscent of the one taken in~\cite{LSNR2013}, where the discretization yields a time-modulated Poisson process, whereas in our case we deal with a binomial one, as indicated by~\eqref{eq:b}.
In order to take into account the non-stationarity of Beta arrivals, we adapt \eqref{eq:rl_total} in the following way
 \begin{align}
\label{eq:rl_total_beta}
    	r^*( \parresource^*, \latency ) = \sum_{t_s=1}^{T_A} \sum_{\nactiveuser} r ( \parresource^*, \latency, \nactiveuser ) {\frac{\Pr [ \nactiveuserrv = \nactiveuser | t_s ]}{ T_A / \latency}} \geq \reliability,
 \end{align}
 \vspace{-0.1cm}
where $\frac{1}{ T_A / \latency }$ is the probability to select any of the $T_A / \latency$ intervals, used according to the law of total probability to calculate the expected number of \device s per interval.
Finally, in the rest of the text, we will assume that duration of $\latency$ is equal to 10~ms for the Beta arrivals.  

\subsection{Comparison of analysis and simulations}

In order to validate the analysis, we have implemented a discrete-time Monte Carlo simulator in MATLAB.
We used Poisson arrivals and Beta arrivals with matching expected number of arrived users in a batch (i.e., average access load), varying $\mpr$, fixing $\text{L}=5$ and $\reliability = 0.99$.\footnote{We recall that duration of $\latency$ is assumed 10~ms for Beta arrivals, which is equal to 5 generic time units in this section.}
For the sake of precision, we have run $10^6$ iterations for each scenario.
At the start of each simulation run, the algorithm is provided $g \, \in \,\{1, \cdots, 40\}$ channels to select the minimum from, using \eqref{eq:gopt} and \eqref{eq:gstar}; the number of channels is limited to demonstrate a realistic scenario.
{If no value of $g$ is able to fulfill the reliability-latency constraint {for the given $\mpr$}, the algorithm outputs $\parresource=0$ and that scenario is not simulated.}

Fig.~\ref{fig:gstar_rel_sims} shows simulated $r^*(\parresource^*, \latency)$ of FSA with $\mpr$-MPR when $\parresource^*$ is chosen according to \eqref{eq:gstar}, for Poisson and Beta arrivals and $\reliability = 0.99$ (note that $\mpr=1$ can not fulfill the requirements in the considered scenario).
The results reveal that, as $\mpr$ increases, $r^*(\parresource^*, \latency)$ becomes larger than the requirement $\reliability$.
This is due to the fact that with increasing $\mpr$, i.e., increasing size of $\mpr$-superslots, reflects in the granularity of the choice in $\parresource^*$ (recall that $\latency$ is fixed), and consequentially in the granularity of potential values of $r^*(\parresource^*, \latency)$.
This overshooting of the reliability requirement also influences the throughput performance, as discussed next.

Fig.~\ref{fig:gstar_tpt_sims} shows throughput performance for the same settings as in Fig.~\ref{fig:gstar_rel_sims}.
The circles and pluses denote the simulation and analytical results, respectively; obviously, the results match. 
It can be seen that higher throughputs can be achieved for Poisson arrivals, which can be expected due to the bursty behavior of Beta arrivals.
Further, increasing $\mpr$ benefits the throughput in general.
However, depending on the interplay between the values of the average load, $\mpr$, $\reliability$ and $L$, it may turn out that throughput drops after $\mpr$ exceeds some value; this is shown in Fig.~\ref{fig:gstar_tpt_sims}(a), where the optimal $K$ is 5 and not 10, which reflects the identified overshooting of the reliability-latency requirement shown in Fig.~\ref{fig:gstar_rel_sims}.
The similar effect also exists in Figs.~\ref{fig:gstar_vs_g_tpt_pois} and \ref{fig:gstar_vs_g_tpt_beta}.

%

\subsection{Comparison of $\mathrm{T}$ and $\mathrm{T}^*$}

We now compare throughputs given by \eqref{eq:t_g} and \eqref{eq:t_g*}, in order to provide insights of the role of the knowledge on the number of arrived users.
We assume that $\reliability  \in \{0.99, 0.99999\}$, $\latency = 5$, and investigate throughput performance for Poisson arrivals with $\lambda \in \{ 3, 15\}$ and Beta arrivals with $n_\text{tot} \in \{ 3000, 15000\}$.  

\begin{figure*}[t!]
 	\centering
 	\begin{subfigure}{0.47\textwidth}
 		\includegraphics[width = 1\textwidth]{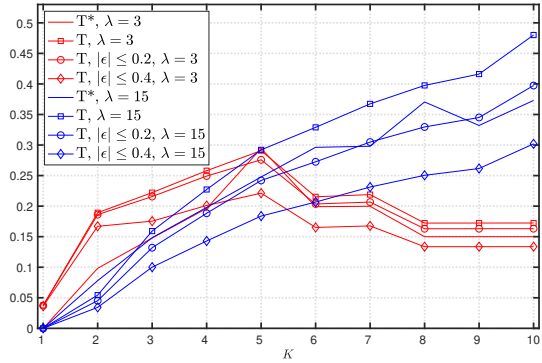}
 		\caption{$R=0.99$ }
 		\label{fig:gstar_g_09}
 	\end{subfigure}
 	\begin{subfigure}{0.47\textwidth}
 		\includegraphics[width = 1\textwidth]{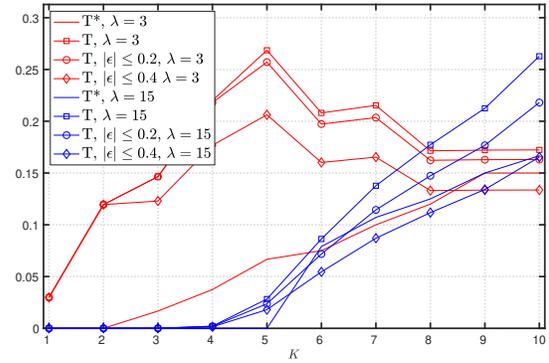}
 		\caption{$R=0.99999$  }
 		\label{fig:gstar_g_099}
 	\end{subfigure}
 	\caption{Comparison of $\mathrm{T}$ and $\mathrm{T}^*$ for for different reliability constraints as function of $K$, $L=5$, Poisson arrivals.}
 	\label{fig:gstar_vs_g_tpt_pois}
 \end{figure*}

Fig.~\ref{fig:gstar_vs_g_tpt_pois} is dedicated to Poisson arrivals and shows that $\mathrm{T}$ outperforms $\mathrm{T}^*$, as it could be expected. This effect is more pronounced for the higher value of $\reliability$.
Since the number of arrived users is typically not known a priori, but has to be estimated, we also investigated the impact of the potential estimation error on the throughput performance.
Specifically, we assume that the relative estimation error $\epsilon$ is bound as $| \epsilon | \leq \epsilon_\text{max}$, and that the algorithm for selection of the number of frequency channels in \eqref{eq:gopt} performs over-provision by assigning $\parresource_\text{o} (\nactiveuser) = g(\lceil n_a \cdot  (1+\epsilon_\text{max}) \rceil)$ frequency channels.
The impact of the estimation error (and the related over-provisioning) on $\mathrm{T}$ is also depicted in Fig.~\ref{fig:gstar_vs_g_tpt_pois}, assuming that $\epsilon_\text{max} = \{0.2, 0.4 \}$, which may be considered as quite high values.
Obviously, the over-provisioning plays it's role by decreasing $\mathrm{T}$, such that for $\reliability = 0.99$,  $\mathrm{T}$ becomes similar or worse than $\mathrm{T}^*$ 
Nevertheless, for higher $\reliability$, $\mathrm{T}$ with over-provisioning may significantly outperform $\mathrm{T}^*$, as shown in Fig.~\ref{fig:gstar_vs_g_tpt_pois}(b).

\begin{figure*}[t!]
 	\centering
 	\begin{subfigure}{0.47\textwidth}
 		\includegraphics[width = 1\textwidth]{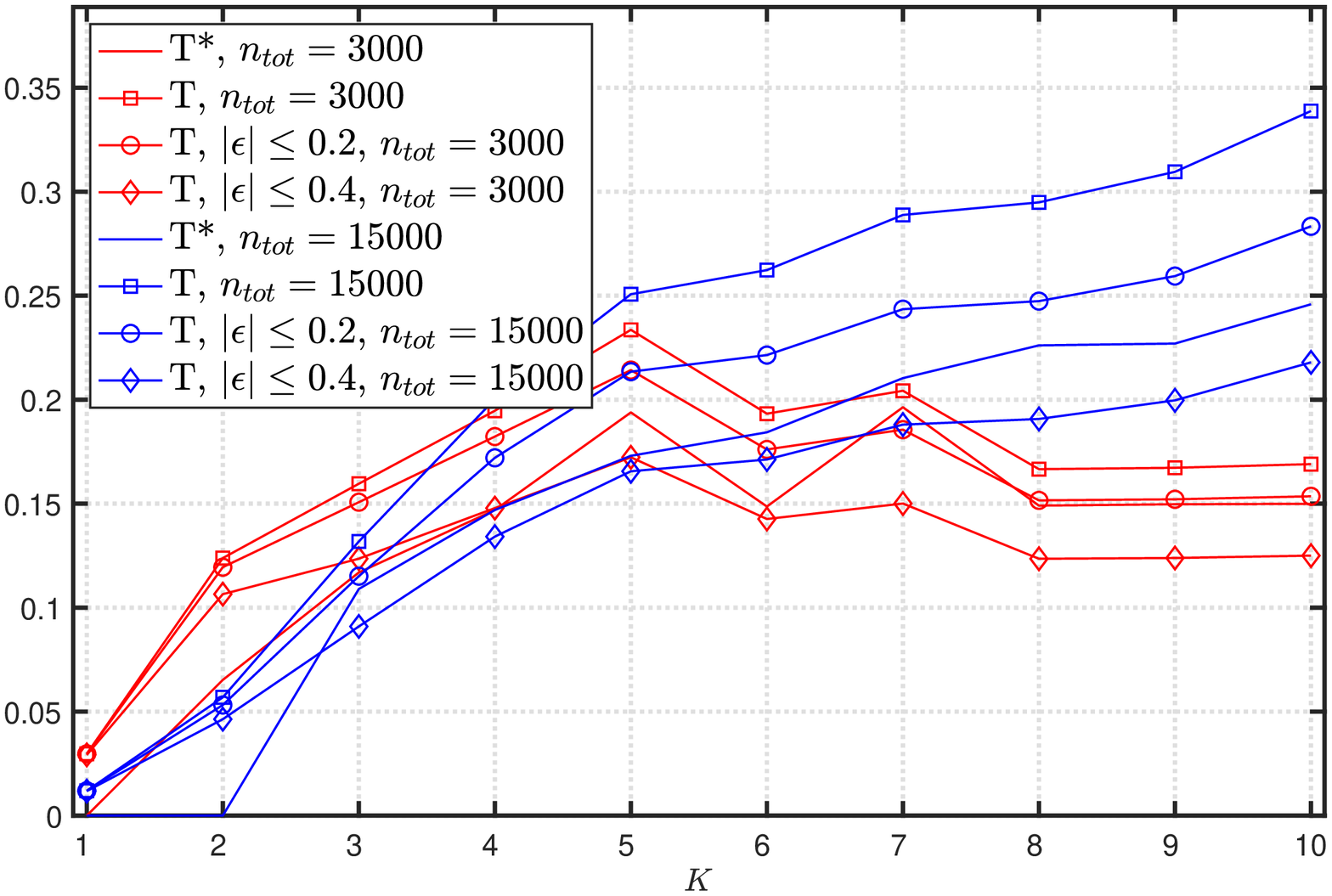}
 		\caption{$R=0.99$ }
 		\label{fig:gstar_g_09b}
 	\end{subfigure}
 	\begin{subfigure}{0.47\textwidth}
 		\includegraphics[width = 1\textwidth]{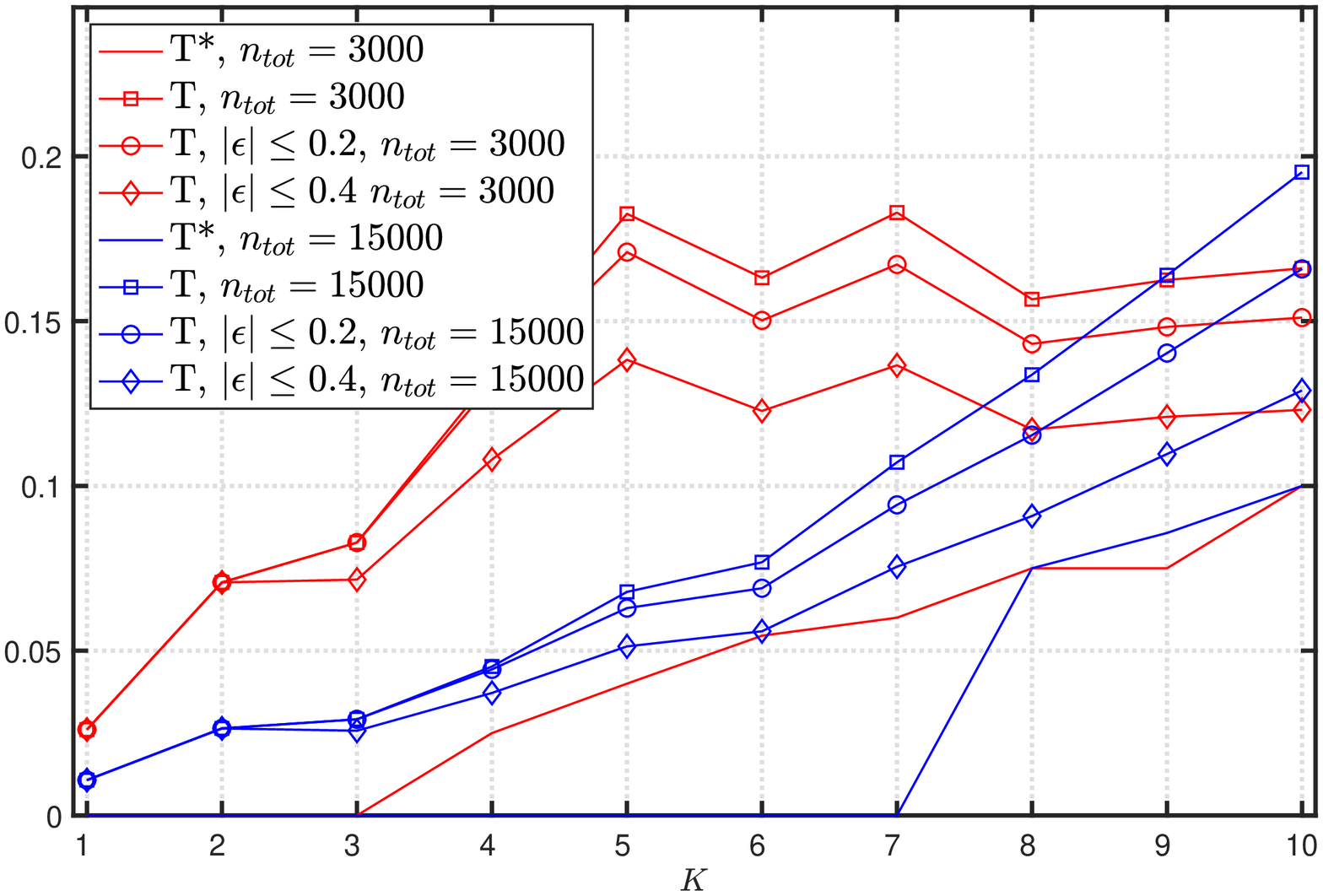}
 		\caption{$R=0.99999$  }
 		\label{fig:gstar_g_099b}
 	\end{subfigure}
 	\caption{Comparison of $\mathrm{T}$ and $\mathrm{T}^*$ for different reliability constraints as function of $K$, $\latency=5$, Beta arrivals.}
 	\label{fig:gstar_vs_g_tpt_beta}
 	\vspace{-0.1cm}
 \end{figure*}

Fig.~\ref{fig:gstar_vs_g_tpt_beta} corresponds to the case of Beta arrivals, showing that, in comparison to Fig.~\ref{fig:gstar_vs_g_tpt_pois}, the gains in performance of $\mathrm{T}$ are more pronounced.
For instance, $\mathrm{T}$ with $\epsilon_\text{max} = 0.2$ fares better than $\mathrm{T}^*$ when $\reliability = 0.99$.
When $\reliability = 0.99999$, $\mathrm{T}$ is better than $\mathrm{T}^*$  even with $\epsilon_\text{max} = 0.4$. 
We also note that both Fig.~\ref{fig:gstar_vs_g_tpt_pois}  and Fig.~\ref{fig:gstar_vs_g_tpt_beta} show that estimation is able to ``unlock'' the use of lower values of $\mpr$, with respect to case when only the knowledge of the arrival distribution is used to dimension $\parresource$.
For instance, in Fig.~\ref{fig:gstar_vs_g_tpt_beta}(b), one cannot have $\mpr=1$ in the latter case, as $\reliability$ constraint cannot be satisfied and the throughout is 0.
On the other hand, $\mpr=1$ can be used if the estimation of the number of arrived users is performed, even with a high relative error.

In Tab.~\ref{tab:fsa} we shared the values of the ratio $\frac{\mathrm{T} -\mathrm{T}^*}{\mathrm{T}^*}$, which could be understood as the measure of the normalized gain in throughput if the number of arrived users is known.
We see that the gain increases with increasing $\reliability$ and increasing average load.
Also, the gain for Beta arrivals is higher, as the distribution has a higher variance compared to Poisson arrivals. 

\begin{figure*}[t!]
  	\centering
  	\begin{subfigure}{0.49\textwidth}
  		\includegraphics[width = 1\textwidth]{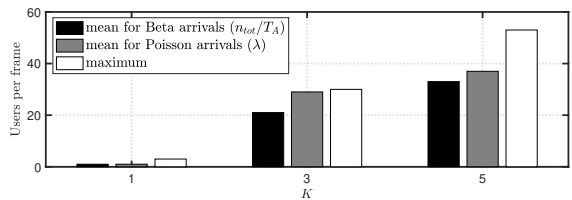}
  		\caption{$R=0.99$ }
  		\label{fig:comp_user_099}
  	\end{subfigure}
  	\begin{subfigure}{0.49\textwidth}
  		\includegraphics[width = 1\textwidth]{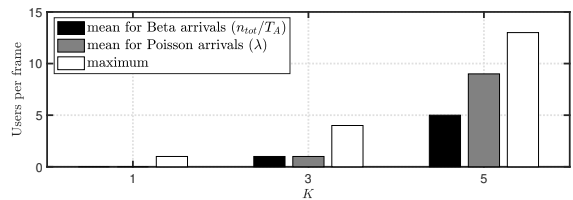}
  		\caption{$R=0.99999$  }
  		\label{fig:comp_user_09999}
  	\end{subfigure}
  	\caption{Number of users supported per frame for different reliability constraints as function of $K$, $g=40$, $\latency=5$. }
  	\label{fig:comp_user}
  	 	\vspace{-0.1cm} 
  	 		\vspace{-0.1cm}
  \end{figure*}

Finally, in order to additionally illustrate the benefits of the knowledge of the number of arrived users, we consider a scenario in which the number of resources, i.e., $\latency$ and $\parresource$ are fixed to $\latency=5$ and $\parresource = 40$.
Further, we assume that $\reliability \in \{ 0.99, 0.99999 \}$, and investigate for FSA with $\mpr$ the following: (i) what is the maximum number of users that can be admitted in the system at any given moment, (ii) average number of users that can be admitted in case of Poisson arrivals, and (iii) average number of users that can be admitted in case of Beta arrivals, 
Fig.~\ref{fig:comp_user} shows the corresponding results.
Obviously, with increasing $\mpr$, the gap between Poisson and Beta arrivals increases (i.e., $\lambda$ becomes increasingly larger than $\frac{ \totaluser}{{T_A}/\latency} $), and the trend holds for the maximum number of users versus the Poisson and Beta arrivals.
This implies that, if one is able to estimate the actual number of arrived users, then one could use the existing resources much better, by letting in the system more users besides the ones that belong to the Poisson/Beta arrival process. 

\vspace{-0.1cm}
\vspace{-0.1cm}
\section{Discussion and Conclusions}
\label{sec:conc}
\begin{table}[t!]
	\centering
	\vspace{0.4cm}
	\begin{tabular}{|c|c|c|c|c|c|}
		\hline
		Scenario	& Arrival	 & $K=1$ & $K=3$   & $K=5$   & $K=10$ \\
		\hline
		{R=0.99} &
		$\lambda=3$&    0.0374 & 0.0743&	0.0036&		0.0224 \\	\cline{2-6} 
		
		&$\lambda=15$&    $\infty$ & 0.0731 & 0.1793&		0.2870 \\		\cline{2-6}  
		& $\totaluser =3000$&    $\infty$ & 0.3650 &	0.2049 &		0.1274 \\		\cline{2-6}	
		& $\totaluser =15000$&    $\infty$ & 0.2095 & 0.4491&		0.3779 \\	\hline	 
		{R=0.99999}&  $\lambda=3$ &       $\infty$	& 7.7919 &	3.0286 &	0.1488 \\ 	\cline{2-6} 
		& $\lambda=15$ &       $\infty$	& $\infty$ &	$\infty$  &	0.5767 \\ \cline{2-6}
		&	 $\totaluser =3000$ &       $\infty$	&  $\infty$ &	3.5643 & 0.6601 \\ 	\cline{2-6} 
		\cline{2-6}  
		& $\totaluser =15000$ &       $\infty$	& $\infty$ &	$\infty$  &	0.9521 \\ \hline
	\end{tabular}
	\caption{Normalized throughput gain $\frac{\mathrm{T} -\mathrm{T}^*}{\mathrm{T}^*}$. The cases that are infeasible for \eqref{eq:rl_total} and feasible for  \eqref{eq:rl}, are denoted by $\infty$.}
	\label{tab:fsa}
	 	\vspace{-0.1cm}
	 	 	\vspace{-0.1cm}
\end{table}
In this paper, we have evaluated grant-free access scheme with reliability and latency constraints.
We based our analysis on framed slotted ALOHA with $K$-MPR.
FSA represents a single shot transmission algorithm, while $K$-MPR is an abstraction of non-orthogonal multiple access schemes, seen as potential multiplexing solution in the coming 5G systems.

To stress the radio resource management perspective, we provide definition of throughput for reliability-latency constrained grant-free access. 
The throughput is evaluated given different information on user activity i.e., the knowledge of the actual number of arrived users or just the knowledge of the arrival distribution.
We have shown that with increasing reliability-latency requirements, the knowledge of the number of arrived users becomes more beneficial for the throughput performance.

However, the information about the number of arrived users is typically not readily available in grant-free access, but could be obtained using an estimation algorithm.
In turn, the estimation algorithm always involves an estimation error and also requires time-frequency resources for its execution, where it is reasonable to assume that the estimation error will decrease as the number of resources dedicated to the estimation increase.
We investigated the effects of the former in the paper, by introducing over-provisioning of the resources for the grant-free access, that will counteract the potential estimation error.
We showed that for high reliability levels, throughput performance still benefits from the estimation even for the considerable estimation error levels.
In this regard, we note that the 3GPP model foresees higher average access loads in the cell~\cite{3rdGenerationPartnershipProject3GPP,LSNR2013} than the ones assumed in the paper, and we conjecture that the throughput gains in this case will be even more pronounced.
Finally, we note that the investigation of the impact of the amount of resources used for the estimation on the throughput performance, including the trade-offs between the estimation accuracy and the amount of resources used for the estimation, is left for future work.

 	\vspace{-0.1cm}
 	 	\vspace{-0.1cm}

\appendix

\subsection*{Reliability of FSA with K-MPR}
\label{app:2}

We start with the basic definition given in \eqref{eq:rl}
 	\vspace{-0.1cm}
\begin{equation}
	r ( \parresource, \latency, \nactiveuser) = \frac{\expectation[\numberresolved]}{\nactiveuser}.
	\label{eq:define_rel}
\end{equation}
 	\vspace{-0.1cm}
For FSA with $\mpr$-MPR, we can calculate $\expectation[\numberresolved]$ as the product of the expected number of resolved users in a $K$-superslot and the number of $K$-superslots 
 	\vspace{-0.1cm}
\begin{equation}
	\expectation[ \numberresolved ] = \left \lfloor \frac{\parresource\latency}{\mpr} \right \rfloor \expectation[\nsucuserresource]
	\label{eq:define_rel_2}
\end{equation}
where $\nsucuserresource$ is the random variable denoting the number of resolved users in a $\mpr$-superslot.

The expected number of resolved users in a $\mpr$-superslot can be calculated in the following way
\begin{equation}
	\expectation[\nsucuserresource]=
	\begin{cases}
		\sum_{i=1}^{\mpr} i \, {{\nactiveuser}\choose{i}} \simprob^i\left(1-\simprob\right)^{\nactiveuser-i} & \text{if } \nactiveuser\geq \mpr    \\
		\simprob \, \nactiveuser  & \text{else}  \text{,}	
	\end{cases}
	\label{eq:expected_succ_perres}
\end{equation}
where $\simprob$ is the $\mpr$-superslot selection probability by an arrived user, given by
$\simprob = \frac{1}{\lfloor\frac{\parresource\latency}{\mpr}\rfloor}$.
We simplify the expectation further for $\nactiveuser \geq K$ as
\begin{equation}
	\expectation[\nsucuserresource]=\simprob \, \nactiveuser \left({1-\simprob}\right)^{\nactiveuser-1}	\sum_{i=0}^{\mpr-1} {{\nactiveuser-1}\choose{i}} \left(\frac{\simprob}{1-\simprob}\right)^{i}\text{.}
	\label{eq:expected_succ_perres6}
\end{equation}
Finally, by plugging \eqref{eq:expected_succ_perres6} into \eqref{eq:define_rel_2} and then into~\eqref{eq:define_rel}, we get
\begin{equation}
	r ( \parresource, \latency, \nactiveuser) = \left({1-\simprob}\right)^{\nactiveuser-1}	\sum_{i=0}^{\mpr-1} {{\nactiveuser-1}\choose{i}} \left(\frac{\simprob}{1-\simprob}\right)^{i}.
	\label{eq:def_rel_mcsa_mpr3}
\end{equation}

\section*{Acknowledgement}

The work has partly been supported by the ERC Conso-lidator Grant nr. 648382 WILLOW and partly by the Horizon 2020 project ONE5G (ICT- 760809).
The views expressed in this contribution are those of the authors.

\bibliographystyle{IEEEtran}
\bibliography{main}

\end{document}